\documentclass[11pt,journal,twocolumn, romanappendices]{IEEEtran}
\usepackage{cite,calc,color}
\usepackage[cmex10]{amsmath} 
\usepackage{amsfonts,amsthm,amssymb,graphics,subfigure,epsfig,graphics,mathrsfs}
\usepackage[mathcal]{euscript} 
\usepackage[T1]{fontenc} 

\newtheorem{thm}{Theorem}

\newtheorem{fact}{Fact}

\newtheorem{rem}{Remark}

\theoremstyle{definition}
\newtheorem{defs}{Definition}

\DeclareMathOperator{\poly}{poly}

\newcommand\cX{\mathcal{X}}
\newcommand\cY{\mathcal{Y}}
\newcommand{\cA}{\mathcal{A}}
\newcommand{\cB}{\mathcal{B}}
\newcommand{\cC}{\mathcal{C}}

\newcommand{\cO}{{\mathcal{O}}}
\newcommand{\cE}{\mathcal{E}}

\newcommand{\cI}{\mathcal{I}}

\newcommand{\cP}{\mathcal{P}}
\newcommand{\cS}{\mathcal{S}}

\newcommand{\cT}{\mathcal{T}}

\newcommand{\ck}{\text{\it k}}
\newcommand{\cK}{\text{\it K}}

\newcommand{\defeq}{\overset{\text{def}}{=}}
\newcommand{\bm}[1]{\mbox{\boldmath{$#1$}}}

\newcommand{\pr}{{\mathbb{P}}}
\newcommand{\ex}{{\mathbb{E}}}
\newcommand{\E}{\cE}
\newcommand{\openone}{\leavevmode\hbox{\small1\normalsize\kern-.33em1}}

\newcommand{\f}{\rho}
\newcommand{\ie}{{\it{i.e.}}}

\begin{document}


\title{Energy and Sampling Constrained \\ Asynchronous Communication}

\author{Aslan Tchamkerten, Venkat Chandar, and Giuseppe Caire
\thanks{This work was supported in part by an Excellence Chair Grant
from the French National Research Agency (ACE
project). } \thanks{A.~Tchamkerten is with the
Department of Communications and Electronics, Telecom
ParisTech, 75634 Paris Cedex 13, France.
Email: aslan.tchamkerten@telecom-paristech.fr.}  \thanks{V.~Chandar
is with MIT Lincoln Laboratory, Lexington, MA 02420, USA. Email:
vchandar@mit.edu.}  
\thanks{G. Caire is with the Viterbi School of
Engineering, University of Southern California,
Los Angeles, USA.
Email: caire@usc.edu.}   }

\maketitle

{\begin{abstract} The minimum energy, and, more generally, the minimum cost, to transmit one bit of information has been recently derived for bursty
communication when information is available infrequently at random times at the
transmitter. This result assumes that the receiver is always in the listening
mode and  samples all channel outputs until it makes
a decision. If the receiver
is constrained to sample only a fraction $\f\in (0,1]$ of
the channel outputs, what is the cost penalty due to sparse output sampling? 

Remarkably, there is no penalty: {\emph{regardless}} of $\f>0$ the asynchronous capacity per unit
cost is the same as under full sampling, \ie, when $\f=1$. Moreover, there is not even a penalty in terms of decoding delay---the elapsed time between when information is available until when it is decoded. 
This latter result relies on the possibility to sample adaptively; the next sample can be chosen as a function of past samples. Under non-adaptive sampling, it is possible to achieve the full sampling asynchronous capacity per unit cost, but the decoding delay gets multiplied by $1/\f$. Therefore adaptive sampling strategies are of particular interest in the very sparse sampling regime.
\end{abstract}}


\normalsize
\section{Introduction}
\label{intro}

{ \IEEEPARstart{I}{n} many emerging technologies, communication is sparse and
asynchronous, but it is essential that when data is available, it is delivered to the destination
as timely and reliably as possible. 
Examples are sensor networks monitoring rare but critical events, such as
earthquakes, forest fires, or epileptic seizures.}

For such settings, \cite{6397617} characterized the
asynchronous capacity per unit cost based on the following
model. There are $B$ bits of information that are made available to the transmitter at some random time $\nu$, and need to be
communicated to the receiver. The $B$ bits are coded and
transmitted over a memoryless channel using a sequence of symbols that have
costs associated with them. The rate $\bm{R}$ per unit cost is the total number
of bits divided by the cost of the transmitted sequence.
Asynchronism is captured here by the fact that the random time $\nu$ is 
not known {\em a priori} to the receiver. However both transmitter and receiver know that $\nu$ is distributed ({\it{e.g.}}, uniformly) over a time horizon 
$[1, \ldots, A]$. At all times before and after the actual transmission, the receiver observes ``pure noise.''
The noise distribution corresponds to a special input ``idle symbol'' $\star$
being sent across the channel (for example, in the case of  a Gaussian channel, this would be the $0$, \ie, no transmit signal). 

The goal of the receiver is to reliably decode the information bits by
sequentially observing the outputs of the channel.

A main result in \cite{6397617} is a single-letter characterization of
the asynchronous capacity per unit cost
$\bm{C}(\beta)$ where $$\beta\defeq \frac{\log A}{B}$$ denotes the {\emph{timing uncertainty per information bit}}. While this result holds for arbitrary discrete memoryless channels and
arbitrary input costs, the underlying model assumes that the receiver 
is always in the listening mode: every channel output is observed until decoding happens. 

What happens when the receiver is constrained to observe a fraction $0<\rho\leq 1$
of the channel outputs?
In this paper, it is shown that the asynchronous capacity per unit
cost is {\emph{not}} impacted by a sparse output sampling. 
More specifically, the asynchronous capacity per unit cost satisfies 
$$\bm{C}(\beta,\f)=\bm{C}(\beta,1)$$
for any asynchronism level $\beta>0$ and sampling frequency $0<\f\leq 1$. Moreover, the decoding delay is minimal: the elapsed time between when information starts being sent and when it is decoded is the same as under full sampling. This result uses the possibility for the receiver to sample adaptively: the next sample can be chosen
as a function of past observed samples. In fact, under non-adaptive
sampling, it is still possible to achieve the full sampling asynchronous capacity per unit cost, but the decoding delay gets multiplied by a factor $1/\rho$ or $(1+\rho)/\rho$ depending on whether or not $\star$ can be used for code design. Therefore, adaptive sampling strategies are of particular interest in 
the very sparse regime. 

We end this section with a brief review of studies related to the above communication model. This model was introduced in \cite{chandar2008optimal,tchamkerten2009communication}. Both of these works focused mainly on the {\emph{synchronization threshold}}---the largest level of asynchronism under which it is still possible to communicate reliably.  In \cite{tchamkerten2009communication, 6352910} communication rate is defined with respect to the decoding delay, the expected elapsed time between when information is available and when it is decoded. Capacity upper and lower bounds are established and shown to be tight for certain channels.  In \cite{6352910} it is also shown that so-called training-based schemes, where synchronization and information transmission use separate degrees of freedom,  need not be optimal in particular in the high rate regime. 

The finite message regime has been investigated by Polyanskiy in \cite{6365818} when capacity is defined with respect to the codeword length, \ie, same setting as \cite{6397617} but with unit cost per transmitted symbol. A main result in  \cite{6365818} is that dispersion---a fundamental quantity that relates rate and error probability in the finite block length regime---is unaffected by the lack of synchronization. Whether or not this remains true under sparse output sampling is an interesting open issue.

 Note that the seemingly similar notions of rates investigated in \cite{tchamkerten2009communication,6352910} and \cite{6397617,6365818} are in fact very different. In particular, capacity with respect to the expected decoding delay remains in general an open problem. 
 
A ``slotted'' version of the above communication model  was considered in \cite{wang2011error} by Wang, Chandar, and Wornell where communication now can happen only in one of consecutive slots of the size of a codeword. For this model, the authors investigated the tradeoff between the false-alarm event (the decoder declares a message before even it is sent) and the miss event (the decoder misses the sent codeword).

The previous works consider point-to-point communication. A (diamond) network configuration was recently investigated by Shomorony, Etkin, Parvaresh, and Alvestimehr in \cite{shomorony2012bounds} who provided bounds on the minimum energy needed to convey one bit of information across the network.

In above models, although communication is bursty, information transmission is contiguous since it always lasts the codeword duration. A complementary setup proposed by Khoshnevisan and Laneman \cite{6283479} considers a bursty communication scenario caused by an intermittent codeword transmission. This model can be seen as a slotted variation of the purely insertion channel model, the latter being a particular case of the general insertion, deletion, and substitution channel introduced by Dobrushin  \cite{D}.

This paper is organized as follows. Section~\ref{moper} contains some
background material and extends the model
developed in \cite{6397617} to allow for sparse output sampling. 
Section~\ref{revres} contains the main results and briefly discusses extensions to a decoder-universal setting and to a multiple access setup.
Finally Section~\ref{analysis} is devoted to the proofs.

\section{Model and Performance Criterion}\label{moper}

The asynchronous communication model we consider captures the following general features:
\begin{itemize}
\item Information is available at the transmitter at a random
time;
\item The transmitter can choose when to start sending
information based on when information is available and based on what message needs to be  transmitted;
\item There is a cost associated to each channel input;
\item Outside the information transmission period the
transmitter stays  idle and the receiver observes
noise;
\item The decoder is sampling constrained and can observe only a fraction of the channel outputs.
\item  Without knowing a priori when information is available,
the decoder should decode reliably and as early as possible, on a sequential basis.\end{itemize}

The model is now specified. Communication is discrete-time and carried over a discrete memoryless
channel characterized by its finite input and output alphabets $$\cX\cup \{\star\}\quad \text{and}\quad \cY\,,$$ respectively, and transition probability matrix $$Q(y|x),$$ for
all $y\in \cY$ and $x\in \cX\cup \{\star\}$.  The alphabet $\cX$ may or may not include $\star$. Without loss of generality, we assume
that for all $y\in \cY$ there is some $x\in \cX\cup\{\star\}$ for which $Q(y|x)>0$.

Given $B\geq 1 $ information bits to be transmitted, a codebook ${\cal{C}}$ consists of
$$M= 2^B$$ codewords of length $n\geq 1$ composed of symbols from ${\cal{X}}$. 

\begin{figure}
\begin{center}
\input{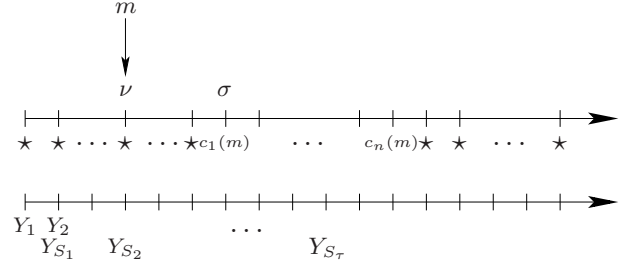}
\caption{\label{grapheesss} Time representation of what is sent (upper arrow)
and what is received (lower arrow). The ``$\star$'' represents the ``idle''
symbol. Message $m$ arrives at time $\nu$ and starts
being sent at time $\sigma$. The receiver samples at
the (random) times $S_1,S_2,\ldots$ and decodes at time
$S_\tau$ based on $\tau$ output samples.}
\end{center}
\end{figure}

A randomly and uniformly chosen message $m$ arrives at the transmitter at a random time $\nu$,
independent of $m$, and uniformly distributed over
$[1,\ldots,A]$, where the integer $$A=2^{\beta B}$$ characterizes the {\emph{asynchronism
level}} between the transmitter and the receiver, and
where the constant $$\beta\geq 0$$ denotes the
{\emph{timing uncertainty per information bit}}, see
Fig.~\ref{grapheesss}.

We consider one-shot communication, \ie, only one message arrives
over the period $[1,2, \ldots, A]\,.$  If $A=1$, the
channel is said to be synchronous.

Given $\nu$ and $m$, the transmitter chooses a time
$\sigma(\nu,m)$ to start sending codeword $c^n(m)\in
\cC$ assigned to message $m$. Transmission cannot
start before the message arrives or after the end of
the uncertainty window, hence  $\sigma(\nu,m)$ must
satisfy $$\nu \leq \sigma (\nu,m)\leq A\quad
\text{almost surely.}$$ In the rest of the paper, we
suppress the arguments $\nu$ and $m$ of $\sigma$ when
these arguments are clear from context.

Before and after the codeword transmission, \ie, before time $\sigma$ and after time $\sigma+n-1$, the receiver
observes ``pure noise,'' Specifically, conditioned on the event $\{\nu=t\}$,
$t\in \{1,\ldots,A\}$, and on the message to be conveyed $m$, the receiver
observes independent channel outputs $$Y_1,Y_2,\ldots,Y_{A+n-1}$$ distributed as
follows. For $$1\leq i\leq \sigma(t,m)-1$$ or $$\sigma(t,m)+n\leq i\leq A+n-1\,,$$
the $Y_i$'s are ``pure noise'' symbols, \ie,$$ Y_i\sim Q(\cdot|\star)\,.$$  For $
\sigma \leq i\leq  \sigma+n-1$
$$Y_i\sim Q(\cdot|{c_{i-\sigma+1}(m)})$$ where $c_{i}(m)$ denotes the $i$\/th symbol
of the codeword $c^n(m)$.

The receiver operates according to a sampling strategy and a sequential decoder. A sampling
strategy consists of ``sampling times'' which are
defined as an ordered collection of random time indices
$${\cS}=\{(S_1,\ldots,S_\ell)\subseteq
\{1,\ldots,A+n-1\}:S_i< S_j,i<j\}$$  where $S_j$ is
interpreted as the $j$\/th sampling time. 

 The sampling
strategy is either non-adaptive or adaptive. It is non-adaptive when the
sampling times given by ${\cS}$ are all known before communication starts, hence $\cS$ is independent
of $Y_1^{A+n-1}$. The strategy is adaptive when the sampling times are function of past observations. This means that $S_1$ is an arbitrary value in $\{1,\ldots, A+n-1\}$, possibly random
but independent of $Y_1^{A+n-1}$ and,  for $j\geq 2$, $$S_j=g_j(\{Y_{S_i}\}_{i<j})  $$
for some (possibly randomized) function $$g_j:
\cY^{j-1}\to \{S_{j-1}+1,\ldots,A+n-1\}\, .$$
Notice that $\ell$, the total number of output samples, may be random
under adaptive sampling, but also under non-adaptive sampling,
since the strategy may be randomized (but still
independent of the channel outputs $Y_1^{A+n-1}$).

Once the sampling strategy is fixed, the receiver decodes by means of a sequential test
$(\tau,\phi)$, where $\tau$, the decision time, is a
stopping time with
respect to the sampled  sequence 
$$Y_{S_1},Y_{S_2},\ldots$$ indicating when decoding
happens,\footnote{Recall that a (deterministic or
randomized) stopping time $\tau$ with respect to a sequence of random variables
$Y_1,Y_2,\ldots$ is a positive, integer-valued, random variable such that the
event $\{\tau=t\}$, conditioned on the
realization of $Y_1,Y_2,\ldots,Y_t$, is
independent of the realization of
$Y_{t+1},Y_{t+2},\ldots$, for all $t\geq 1$.}  and where $\phi$ is the decoding function, \ie, a map
 $$\phi: \cO\to
\{1,2,\ldots,M\}$$
where  
$$\cO\defeq \{Y_{S_1},Y_{S_2},\ldots,Y_{S_\tau}\}$$
is the set of  observed samples. Hence, decoding happens at time
$S_\tau$ on the basis of $\tau$ output samples.
Since there are at
most $A+n-1$ sampling times, $\tau$ is bounded by
$A+n-1$.

A code $(\cC,\cS,(\tau,\phi))$ is defined as a codebook,
a receiver sampling strategy, and a decoder (decision time and decoding function). Throughout the paper, whenever clear from context, we
often refer to a code using the codebook symbol $\cC$
only, leaving out an explicit reference to the sampling strategy and to the decoder.

\begin{defs}[Error probability]
The maximum (over messages) decoding error probability
of a code $\cC$ is defined as
\begin{align}\label{maxerror}
\pr({\EuScript{E}}|\cC)\defeq\max_m\frac{1}{A}\sum_{t=1}^A
\pr_{m,t} ({\EuScript{E}}_m),
\end{align}
where the subscripts ``$m,t$'' denote conditioning on the
event that message $m$ arrives at
time $\nu=t$, and where $\EuScript{E}_m$ denotes the
error event that the decoded message does not
correspond to $m$, \ie,
$$\E_m\defeq\{\phi(\cO)\ne m\}\,.$$
\end{defs}

\begin{defs}[Cost of a Code]\label{costc}
The (maximum) cost of a code ${\cal C}$ with respect to a cost function $\ck: \cX\to [0,\infty]$ is defined as
$$ \cK({\cal C})\defeq \max_m \sum_{i=1}^n \ck(c_i(m)).$$
\end{defs}
\noindent{\it{Assumption:}} throughout the paper we make the assumption that the only possible zero cost symbol is $\star$. When $\star \in \cX$ the transmitter can stay idle at no cost. When $\star \notin \cX$ then $\ck(x)>0$ for any $x\in \cX$, which captures the situation where a ``standby'' mode may not be possible at zero cost. The other cases---investigated in \cite{6397617} under full sampling---are either trivial (when $\cX$ contains two or more zero costs symbols) or arguably unnatural ($\cX$ contains a zero cost symbol that differs from $\star$ or when $\star\in \cX$ and all $\cX$ contains only nonzero cost symbols). 

Below, $\pr_m$ denotes the output distribution conditioned on the sending of
message $m$. Hence, by definition we have
$$\pr_m(\cdot)\defeq \frac{1}{A}\sum_{t=1}^A\pr_{m,t}(\cdot)\,.$$
\begin{defs}[Sampling Frequency of a Code]\label{costc}
Given $\varepsilon>0$, the sampling frequency of a code $\cC$, denoted by
$\rho ({\cal{C}},\varepsilon)$, is the relative number of channel outputs that are observed until a message is
declared. Specifically, it is defined as the smallest $r\geq 0$ such that
 $$\min_m\pr_m({\tau}/{S_\tau}\leq
 r)\geq1-\varepsilon\,.$$
 (Recall that $S_\tau$ refers to the last sampling time.)
\end{defs}

\begin{defs}[Delay of a Code]\label{def:delaiscode}
Given $\varepsilon>0$, the (maximum) delay of a code ${\cal
C}$, denoted by $d({\cal
C}, \varepsilon)$, is defined as the smallest integer $l$ such that
$$\min_m\pr_m(S_\tau-\nu \leq  l-1) \geq
1-\varepsilon\,.$$
\end{defs}

We now define capacity per unit cost under the constraint that the
receiver has  access only to a limited number of channel outputs:

\begin{defs}[Asynchronous Capacity per Unit Cost under Sampling Constraint]
\label{def:cap}
\bm{R} is an achievable rate per unit cost at
timing uncertainty per information bit $\beta$ and sampling frequency $\f$, if
 there exists a sequence of codes $\{{\cal C}_B\}$
 and a sequence of positive numbers
 $\varepsilon_B$ with $\varepsilon_B\overset{B\to \infty}{\longrightarrow}0$
 such that for all $B$ large enough
 \begin{enumerate}
\item $\cC_B$ operates at timing uncertainty per information bit $\beta$;
\item the maximum error probability $\pr(\cE|{\cal{C}}_B)$ is at most
$\varepsilon_B$;
\item the rate per unit cost $$\frac{B}{\cK({\cal C}_B)}$$ is at least $
\bm{R}-\varepsilon_B$;
\item the sampling frequency satisfies
$\rho({\cal{C}}_B,\varepsilon_B)\leq \rho +\varepsilon_B$;
\item the delay satisfies\footnote{Throughout the paper $\log$ is always to the base
$2$.}
 $$ \frac{1}{B}\log(d({\cal
C}_B,\varepsilon_B)) \leq \varepsilon_B\,.$$
\end{enumerate}
Notice that the last requirement asks for a subexponential delay.

The asynchronous capacity per unit cost, denoted by $\bm{C}(\beta,\f)$,
is the supremum of achievable rates per unit cost.
\end{defs}

Two basic observations:
\begin{itemize}
\item
$\bm{C}(\beta,\f)$  is a non-increasing function of $\beta$ for fixed $\f$;
\item $\bm{C}(\beta,\f)$  is an non-decreasing function of $\f$ for fixed
$\beta$.
\end{itemize}
In particular, for any fixed $\beta\geq 0$
$$\max_{\f\geq 0}\bm{C}(\beta,\f)=\bm{C}(\beta,1)\,.$$

Capacity per unit cost under full sampling $\bm{C}(\beta,1)$ is characterized in the following theorem:

\begin{thm}[\cite{6397617} Theorem 1]\label{fullsampling}
For any $\beta\geq 0$
\begin{align}\label{capexpr}
 \bm{C}(\beta,1)=\max_X \min\left\{\frac{I(X;Y)}{\ex[\ck(X)]} ,
\frac{I(X;Y ) + D(Y||Y_\star)}{\ex[\ck(X)](1 + \beta)}\right\},
\end{align}
where $\max_X$ denotes maximization with respect to the channel input distribution $P_X$, where 
$(X,Y) \sim P_X(\cdot) Q(\cdot|\cdot)$,  where $Y_\star$ denotes the random output of the channel when the
idle symbol $\star$ is transmitted (\ie, $Y_\star\sim Q(\cdot|\star)$), where
$I(X;Y)$ denotes the mutual information between $X$ and $Y$, and where $D(Y||Y_\star)$ denotes the divergence
(Kullback-Leibler distance) between the distributions of $Y$ and~$Y_\star$.\footnote{$Y_\star$ can be
interpreted as ``pure noise.''} \hfill \QED
\end{thm}
Let $P_{X^*}$ be a capacity per unit cost achieving input distribution, \ie, $X^*$ achieves the maximum in \eqref{capexpr}. As shown in the converse of the proof of \cite[Theorem 1]{6397617}, codes that achieve the capacity per unit cost can be restricted to codes of (asymptotically) constant composition $P_{X^*}$. Specifically, we have
$$\frac{B}{n_B(P_{X^*}) \ex [k(X^*)]}=  \bm{C}(\beta,1)(1-o(1))\quad (B\to \infty)$$
where $n_B(P_{X^*})$ denotes the length of the $P_{X^*}$-constant composition codes achieving  $\bm{C}(\beta,1)$. 
Now  define
$$n^*_B\defeq \min_{P_{X^*}}n_B(P_{X^*})= \min_{{X}\in {\cal{P}}} \frac{B}{\bm{C}(\beta,1) \ex [k(X)]}$$
where $${\cal{P}}\defeq\{X:X\:\:\text{achieves the maximum in}\:(2)\}\,.$$
From the achievability and converse of \cite[Theorem 1]{6397617}, $\{n^*_B\}$ represent the smallest achievable delays for codes $\{\bm{C}_B\}$ achieving the asynchronous capacity per unit cost under {\it{full sampling}} $\bm{C}(\beta,1)$ in the sense that 
$$d(\cC_B,\varepsilon_B)\geq n^*_B(1-o(1)) \quad (B\to \infty)$$
for any $\varepsilon_B\to 0$ as $B\to \infty$. 

Our results, stated in the next section, say that the capacity per unit cost under sampling frequency $0<\rho<1$ is the same as under full sampling, \ie, $\rho=1$. To achieve this, non-adaptive sampling is sufficient. However, if we also want to achieve minimum delay, then adaptive sampling is necessary. In fact, non-adaptive sampling strategies that achieve  capacity per unit cost have a delay that grows at least as $$\frac{n^*_B}{\rho}$$ or $$\frac{n^*_B(1+\rho)}{\rho}$$ depending on whether or not $\star\in \cX$.

We end this section with a few notational conventions. We use $\cP^\cX$ to denote the set of
distributions over the finite alphabets $\cX$.
Recall that the type of a string
$x^n\in \cX^n$, denoted by $\hat{P}_{x^n}$, is the probability over $\cX$ that
assigns to each $a\in \cX$ the number of occurrences of $a$ within $x^n$
divided by $n$~\cite[Chapter $1.2$]{CK}. For instance, if $x^3=(0,1,0)$, then
$\hat{P}_{x^3}(0)=2/3$ and $\hat{P}_{x^3}(1)=1/3$. The joint type
$\hat{P}_{x^n,y^n}$induced by a pair of strings $x^n\in \cX^n, y^n\in \cY^n$ is
defined similarly. The set of strings of length $n$ that have type $P$ is
denoted by $\cT_P^n$. The set of all types over $\cX$ of strings of length $n$ is
denoted by $\cP_n^{\cX}$. Finally, we use $\poly(\cdot)$ to denote a function that does not grow
or decay faster than polynomially in its argument. 

Throughout the paper we use the standard ``big-O'' Landau notation to characterize 
growth rates  (see, e.g., \cite[Chapter~3]{CLRS}). 


\section{Results}\label{revres}

In the sequel we denote by $\bm{C}_{\text{a}}(\beta,
\rho)$ and
$\bm{C}_{\text{na}}(\beta,\rho)$ the capacity
per unit cost when restricted to adaptive and non-adaptive sampling, respectively.

Our first result characterizes the capacity per unit cost under
non-adaptive sampling. 
\begin{thm}[Non-adaptive sampling] \label{thm:nonadaptive}
Under non-adaptive sampling it is possible to achieve the full-sampling capacity per unit cost, \ie
\begin{equation*}
 \bm{C}_{\text{na}}(\beta,\f) =\bm{C}(\beta,1) \quad \text{for any }\: \beta>0,\rho>0\,.
\end{equation*}
Furthermore codes $\{\cC_B\}$ that achieve rate $\gamma \bm{C}(\beta, 1)$, $0\leq \gamma \leq 1$, satisfy
$$\lim_{\gamma\to 1}\liminf_{B\to \infty}\frac{d(\cC_B,\varepsilon_B)}{n^*_B}\geq \frac{1}{\rho}$$
when $\star \in \cX$, and satisfy
$$\lim_{\gamma\to 1}\liminf_{B\to \infty}\frac{d(\cC_B,\varepsilon_B)}{n^*_B}\geq \frac{1+\rho}{\rho}$$
 when $\star \notin \cX$. Finally, the above delay bounds are tight: for any $\varepsilon>0$ and $\gamma$ close enough to $1$ there exists $\{\cC_B\}$ and $\varepsilon_B\to 0$ as $B\to \infty$ such that $$\liminf_{B\to \infty}\frac{d(\cC_B,\varepsilon_B)}{n^*_B}\leq \frac{1}{\rho}+\varepsilon$$ for the case $\star \in \cX$, and similarly for the case $\star \notin \cX$.
\end{thm}
Hence, even with a negligible fraction of the channel outputs it is possible to achieve the full-sampling capacity per unit. However, this comes at the expense of delay which gets multiplied by a factor $1/\rho$ or $(1+\rho)/\rho$ depending on whether or not $\star$ can be used for code design. This disadvantage is overcome by adaptive sampling:
\begin{thm}[Adaptive sampling] \label{thm:adaptive}
Under adaptive sampling it is possible to achieve the full-sampling capacity per unit cost, \ie
\begin{equation*}
 \bm{C}_{\text{na}}(\beta,\f) =\bm{C}(\beta,1) \quad \text{for any }\: \beta>0,\rho>0\,.
\end{equation*}
Moreover, there exists $\{\cC_B\}$ and $\varepsilon_B\to 0$ as $B\to \infty$ such that  
$$d(\cC_B,\varepsilon_B)={n^*_B}(1+o(1)).$$
\end{thm}
The first part of Theorem~\ref{thm:adaptive} immediately follows from the first part of Theorem~\ref{thm:nonadaptive} since the set of adaptive sampling strategies include the set of non-adaptive sampling strategies. The interesting part of Theorem~\ref{thm:adaptive} is that adaptive sampling strategies guarantee minimal delay {\emph{regardless}} of the sampling rate $\rho$, as long as it is non-zero.

What is a an optimal adaptive sampling strategy? Intuitively, such a
strategy should sample sparsely, with a sampling frequency of no more
than $\f$, under pure noise---for otherwise the sampling constraint
is violated. It should also sample the entire sent
codeword, and so densely sample during message
transmission---for otherwise a rate per unit cost penalty is
incurred. The main characteristic of a good adaptive sampling strategy is the
criterion under which the sampling mode switches from sparse to dense.
If the criterion is too conservative, \ie, if the probability of
switching under pure noise is too high, we might sample only
part of the codeword, thereby incurring a cost loss. By contrast, if
this probability is too low, we might not be able to accommodate the
desired sampling frequency. 

The proposed  asymptotically optimal sampling/decoding strategy operates as follows---details are deferred to the proof of Theorem~\ref{thm:adaptive}.
 
The strategy starts in the 
sparse mode, taking samples at times $S_j = \lceil j/\f\rceil$, $j=1,2,\ldots$. 
At each $S_j$, the receiver computes the empirical distribution (or type) of the last $\log (n)$ samples.
If the probability of observing this type under pure noise is greater than $1/n^2$,
the mode is kept unchanged and we repeat this test at the next round $j+1$.
Instead, if it is smaller than ${1}/{n^2}$,
then we switch to the dense sampling mode, taking samples 
continuously for at most $n$ time steps.  
At each of these steps the receiver applies a standard typicality decoding based on the past $n$ output samples. If no codeword is typical with the channel outputs 
after these $n$ times steps, sampling is switched back to the sparse mode. 
As it turns out, the threshold $1/n^2$ can be replaced by any decreasing function of $n$ that decreases at least as fast as $1/n^2$ but not faster than polynomially in $n$.

We end this section by considering the specific case when $\beta=0$, \ie, when the channel is synchronous. For a given sampling frequency
$\f$, the receiver gets to see only a fraction $\f$ of the transmitted
codeword (whether sampling is adaptive or non-adaptive) and hence $$\bm{C}(0,\f)=\f\,\bm{C}(0,1)$$ for any
$\f\geq 0$.

How is it possible that sparse output sampling induces a  rate per unit cost loss for synchronous communication
($\beta=0$), but not for asynchronous
communication ($\beta>0$) as we saw in Theorems~\ref{thm:nonadaptive} and \ref{thm:adaptive}? The reason for this is that when
$\beta>0$, the level of asynchronism is exponential in $B$.
Therefore, even if the receiver is constrained to sample only a
fraction $\f$ of the channel outputs, it may still
occasionally sample fully over, say, $\Theta(B)$ channel
outputs, and still satisfy the overall constraint that the
fraction shouldn't exceed $\f$.\footnote{If over a long trip
we have a high-mileage drive, we can still push the car a few times
without impacting the overall mileage.}

\begin{rem}\label{rem1}
Theorems~\ref{thm:nonadaptive} and \ref{thm:adaptive} remain valid under universal decoding, {\it{i.e.}}, the only element from the channel that the decoder needs to know is its output alphabet $\cal{Y}$. This is briefly discussed at the end of Section~\ref{analysis}.
 \end{rem}

\begin{rem}
Consider a multiple access generalization of the point-to-point setting where, instead of one
transmitter, there are 
$$U=2^{\upsilon B}$$ transmitters who
communicate to a common receiver, where $\upsilon$, $0\leq
\upsilon\leq \beta$, denotes the {\it{occupation}} parameter of the
channel. The messages arrival times $\{\nu_1,\nu_2,\ldots,\nu_U\}$ at the transmitters
are jointly independent and uniformly distributed over
$[1, \ldots,A]$ with $A = 2^{\beta B}$ as before. Communication takes place
as in the previous point-to-point case, each user uses
the same codebook,  and
transmissions start at the times
$\{\sigma_1,\sigma_2,\ldots,\sigma_U\}$. Whenever a
user tries to access the channel while it is occupied,
the channel outputs random symbols, independent of the
input (collision model). 

The receiver operates sequentially and declares $U$
messages at the times
$$S_{\tau_1},S_{\tau_1}+S_{\tau_2}, \ldots,
S_{\tau_1}+S_{\tau_2}+\ldots  S_{\tau_U}$$ where
stopping time $\tau_i$, $1\leq i\leq U$, is with
respect to the output samples
$$Y_{S_{\tau_{i-1}}},Y_{S_{\tau_{i-1}+1}},Y_{S_{\tau_{i-1}+2}},\ldots.$$ 

It is easy to check (say, from the Birthday problem \cite{Gr}) that if 
$$\upsilon< \beta /2$$
and hence $U=o(\sqrt{A})=o(2^{\beta B/2})$, the collision
probability goes to zero as $B \to \infty$.
Hence in the regime of large message size,
the transmitters are (essentially) operating orthogonally, and each user can achieve the
point-to-point capacity per unit cost assuming a per/user error probability.
 We may refer to this regime as the regime of ``sparse transmissions,'' relevant in 
a sensor network monitoring independent rare events.

Note that since the users use the same
codebook, the receiver does not know which transmitter
conveys what information. The receiver can only
recognize the set of transmitted messages.

If the receiver is also required to identify the
messages and their transmitters, then each transmitter effectively conveys
$B(1+\upsilon)$ information bits and the capacity per unit cost gets multiplied
by $1/(1+\upsilon)$.

\end{rem}

\section{Analysis}\label{analysis}

The following two standard type results are often used in our
analysis.

\begin{fact}[\hspace{-.01cm}{\cite[Lemma~1.2.2]{CK}}]
\label{fact:1}
\begin{align*}
|\cP_n^{\cX}| &=\poly(n)\,.
\end{align*}
\end{fact}

\begin{fact}[\hspace{-.01cm}{\protect\cite[Lemma 1.2.6]{CK}}]
\label{fact:2}
If $X^n$ is independent and identically distributed (i.i.d.) according
to $P_1\in\cP^\cX$, then 
\begin{equation*}
\poly(n)e^{-nD(P_2\|P_1)}\leq \pr(X^n\in \cT_{P_2}) \leq  e^{-nD(P_2\|P_1)}.
\end{equation*}
for any $P_2\in\cP^\cX_n$.
\end{fact}

\begin{IEEEproof}[Achievability of Theorem~\ref{thm:nonadaptive}]
Fix some arbitrary distribution $P$ on $\cal{X}$. Let $X$ be the input having that distribution
and let $Y$ be the corresponding output, \ie, $(X,Y)\sim P(\cdot)Q(\cdot|\cdot)$.

Given $B$ bits of information to be transmitted, the codebook ${\cal C}$ is randomly
generated as follows.
For each message $m = 1, \ldots, M$, randomly generate
length $n$ sequences $x^n$ i.i.d. according to $P$, until 
$x^n$ belongs to the ``constant composition'' set\footnote{$||\cdot ||$ refers to the $L_1$-norm.}
\begin{align}\label{constprop}
{\cal{A}}_n=\{x^n:||\hat{P}_{x^n}-P||\leq 1/\log n\}\,.
\end{align}
If (\ref{constprop}) is satisfied, then let $c^n(m) = x^n$ and move to the next message. Stop when
a codeword has been assigned to all messages.
From Chebyshev's inequality, for any fixed $m$, no repetition will be required with high probability 
to generate $c^n(m)$, \ie,
\begin{align}\label{typset}
P^n({\cal{A}}_n)\to 1\quad \text{as}\quad n\to \infty
\end{align}
where $P^n$ denotes the order $n$ product distribution of $P$.

The obtained codewords are thus essentially of constant composition---\ie, each symbol appears roughly the same number of times---and have
cost $n\ex[k(X)](1+o(1))$ as $n\rightarrow \infty$ where $k(\cdot)$ is the input
cost function of the channel.

\noindent{\it{Case $\star\in \cX$:}} Information transmission is as follows. For simplicity let us first assume that $1/\rho$ is an integer. 
Codeword symbols can be transmitted only at multiples of $1/\rho$. Times that are integer multiples of $1/\rho$ from now on are referred to as transmission times. 
Given a message $m$ available at time $\nu$, the transmitter sends
the corresponding codeword $c^n(m)$ during the first $n $ information transmission times coming at time $\geq \nu$. In between transmission times the transmitter sends $\star$. Hence, the transmitter sends 
$$c_1(m)\star\ldots\star c_2(m)\star \ldots \star c_3(m) \{\ldots\ldots \} c_n(m)$$
starting at time $\sigma=\sigma(\nu)=\min\{t\geq:\lfloor t/\rho\rfloor\geq \nu\}$.

The receiver operates as
follows.  Sampling is performed only at the transmission times. At transmission time $t$, the decoder computes the empirical distributions $$\hat{P}_{c^n(m), {y}^n}(\cdot,\cdot)$$
induced by the last output samples $y^n$ and all the codewords $\{c^n(m)\}$. If there is a unique message $m$ for which
$$||\hat{P}_{c^n(m),y^n}(\cdot,\cdot) - P(\cdot)Q(\cdot|\cdot)||\leq
2/\log n,$$ 
the decoder stops and declares that message $m$ was sent. 
If two (or more) codewords $c^n(m)$ and $c^n(m')$ relative to
two different messages $m$ and $m'$ are typical with ${y}^n$, the decoder stops and 
declares one of the corresponding messages at random. 
If no codeword is typical with $y^n$,  the decoder repeats the procedure at the next transmission time. If by the time of the last transmission time no message has been declared, the decoder outputs a random message.

We first compute the error probability averaged over codebooks and messages.  
Suppose message $m$ is  transmitted. The error event that the decoder declares
some specific message $m'\ne m$ can be decomposed as\footnote{Notice that the decoder outputs a message with probability one.}
\begin{align}\label{decom}
\{m\to m'\}=\E_1\cup\E_2\,,
\end{align}
where the error
events $\E_1$ and $\E_2$ are defined as
\begin{itemize}
\item $\E_1$: the decoder stops at a time $t$ between $\sigma$ and $\sigma + (2n-2)/\rho$ (including $\sigma$ and $\sigma+(2n-2)/\rho$) and declares $m'$;
\item $\E_2$: the decoder stops
either at a time $t$ before time $\sigma$ or from time $\sigma+(2n-1)/\rho$ onwards and declares $m'$.
\end{itemize}
Note that when event $\E_1$ happens, the observed sequence is generated by the sent codeword. By contrast, when event
$\E_2$ happens, then the observed sequence is generated only by pure noise.

Using analogous arguments as in the achievability of \cite[Proof of Theorem 1]{6397617} we obtain the upper bounds $$\pr_m(\E_1) \leq  2^{-n(I(X;Y)-\varepsilon)}$$
and
$$\pr_m(\E_2) \leq  A\cdot
2^{- n(I(X;Y)+D(Y||Y_\star)-\varepsilon)}$$
which are both valid for any fixed $\varepsilon>0$ provided that $n$ is large enough.

Combining, we get
\begin{align*}\pr_m(m \rightarrow m') \leq &2^{- n(I(X;Y)-\varepsilon)} \\
&+ A \cdot 2^{-n(I(X;Y) +
D(Y||{Y_\star})-\varepsilon)}\,.\end{align*}

Hence, taking a union bound over all possible wrong
messages, we obtain that for all $\varepsilon>0$,
\begin{align}\label{er11}\pr({\E})\leq 2^B&\Big(2^{- n(I(X;Y)-\varepsilon)}
\nonumber\\
&+ A \cdot 2^{-n(I(X;Y) +
D(Y||{Y_\star})-\varepsilon)}\Big)\nonumber\\
&\defeq \varepsilon_1(n)
\end{align}
for $n$  large enough.  

We now show that the delay of our coding scheme in
the sense of Definition~\ref{def:delaiscode} is at most $n/\rho$. Suppose a specific
(non-random) codeword $c^n(m)\in \cal{A}$ is sent.
If $$\tau> \sigma +(n-1)/\rho\,,$$ then necessarily $c^n(m)$ is not typical with ${Y}^{\sigma+(n-1)/\rho}_\sigma$. By Sanov's theorem this happens with vanishing error probability and hence
$$\pr(\tau-\sigma \leq (n-1)/\rho)=1-\varepsilon_2(n)$$
with $\varepsilon_2(n)\to 0$ as $n\to \infty$. Hence, since $\nu\leq \sigma <\nu+1/\rho$, we get
$$\pr(\tau-\nu \leq n/\rho)=1-\varepsilon_2(n)\,.$$

The proof can now be concluded. From inequality  \eqref{er11} there exists a specific code $\cC\subset{\cA}_n$ whose error
probability, averaged over messages, is less than
$\varepsilon_1(n)$.  Removing the half of the codewords with the highest error
probability, we end up with a set ${\cal{C}}'$ of $2^{B-1}$ codewords 
whose maximum error probability $\pr(\E)$ is such that 
\begin{align}\label{mwi}\pr(\E)\leq
2\varepsilon_1(n)\,,
\end{align} and whose delay satisfies
$$d(\cC',\varepsilon_2(n))\leq n/\rho\,.$$ 
Now fix the ratio $B/n$
and substitute $A = 2^{\beta B}$ in the
definition of $\varepsilon_1(n)$ (see \eqref{er11}). Then, $\pr(\E)$ goes to
zero as $B\rightarrow  \infty$ whenever 
\begin{align}
\label{cn1}
\frac{B}{n} <\min\bigg\{I(X;Y),
\frac{I(X;Y)+D(Y||{Y_\star})}{1+\beta}\bigg\}.
\end{align}
Recall that  by construction, all the codewords  have cost $n\ex[k(X)](1+o(1))$ as
$n\rightarrow \infty$. Hence, for any $\eta>0$ and all $n$ large enough
\begin{align}\label{kos}k(\cC')\leq n\ex[k(X)](1+\eta)\,.
\end{align}
Condition \eqref{cn1} is thus implied by 
condition \begin{align}\label{eqmaster}
\frac{B}{\cK({\cal C}')} <\min\bigg\{\frac{I(X;Y)}{(1+\eta)\ex[\ck(X)]},
\frac{I(X;Y)+D(Y||{Y_\star})}{\ex[\ck(X)](1+\eta)(1+\beta)}\bigg\}.
\end{align}
Maximizing over all input distributions and using the
fact that $\eta>0$ is arbitrary proves that $\bm{C}(\beta,1)$---where
$\bm{C}(\beta,1)$ is defined in Theorem~\ref{fullsampling}---is asymptotically 
achieved by non-random codes with delay no larger than $n/\rho$ with probability approaching one as $n\to \infty$.

 Finally, if $1/\rho$ is not an integer, it suffices to define transmission times as $$t_j=\lfloor j/\rho\rfloor.$$ This guarantees the same asymptotic performance as for the case where $1/\rho$ is an integer.

\noindent{\it{Case $\star\notin \cX$:}} Parse the entire sequence $\{1,2,\ldots,A+n-1\}$ into consecutive superperiods of size $n/\rho$---take $\lfloor n/\rho\rfloor$ if $n/\rho$ is not an integer. The periods of duration $n$ occurring at the end of each superperiod are referred to as transmission periods.  Given $\nu$, the codeword starts being sent over the first transmission period starting at a time $> \nu$. In particular, if $\nu$ happens over a transmission period, then the transmitter delays the codeword transmission to the next superperiod.

The receiver sequentially samples only the transmission periods. At the end of a transmission period,  the decoder computes the empirical distributions $$\hat{P}_{c^n(m), {y}^n}(\cdot,\cdot)$$
induced by the last output samples $y^n$ and all the codewords $\{c^n(m)\}$. If there is a unique message $m$ for which
$$||\hat{P}_{c^n(m),y^n}(\cdot,\cdot) - P(\cdot)Q(\cdot|\cdot)||\leq
2/\log n,$$ 
the decoder stops and declares that message $m$ was sent. 
If two (or more) codewords $c^n(m)$ and $c^n(m')$ relative to
two different messages $m$ and $m'$ are typical with ${y}^n$, the decoder stops and 
declares one of the corresponding messages at random. 
If no codeword is typical with $y^n$,  the decoder waits for the next transmission period to occur, samples it, and repeats the decoding procedure. Similarly as for the previous case, if at the end of the last transmission period no message has been declared, the decoder outputs a random message.

Following the same arguments as for the case $\star\in {\cX}$ we deduce that \eqref{eqmaster} also holds in this case and that for the delay we have
$$\pr(\tau-\nu\leq n+n/\rho)=1-\varepsilon_2(n)$$
for some $\varepsilon_2(n)\to 0$ as $n\to \infty$. To see this, note that a superperiod has duration $n/\rho$ and that if $\nu$ happens during a transmission period, then the actual codeword transmission is delayed to the next transmission period.
\end{IEEEproof}

\begin{IEEEproof}[Delay Converse of Theorem~\ref{thm:nonadaptive}] 
We consider the cases $\star\in \cX$ and $\star\notin \cX$ separately.

\noindent{\it{Case $\star\in \cX$:}} 
Pick some arbitrary $0<\rho<1$, $\beta>0$ such that $\cC(\beta,\rho) > 0$, and $0<\gamma<1$. Consider a code $\cC_B$ with length $n_B$ codewords that achieves rate per unit cost $\gamma \cC(\beta,\rho)-\varepsilon_B>0$, maximum error probability at most $ \varepsilon_B$, sampling frequency $\rho({\cal{C}}_B,\varepsilon_B)\leq \rho +\varepsilon_B$, and  delay $d_B=d(\cC_B,\varepsilon_B)$, for some  $\varepsilon_B\overset{B\to \infty}{\longrightarrow}0$. The sampling strategy $\cS$ is supposed to be non-adaptive, and for the moment also non-randomized.

Denote by $\cI_{\gamma}$ the event that the decoder samples at least $\gamma n^*_B$ samples of the sent codeword---recall that $n^*_B$ refers to the minimal codeword length, see Section~\ref{revres}. Then by the converse of the \cite[Theorem 1]{6397617} \begin{align}\label{eka}\pr_m(\cI_{\gamma'})=1-o(1)\quad (B\to \infty)
\end{align}
for any message $m$, where $\gamma'=\gamma'(\gamma)$ satisfies $\gamma'=\gamma'(\gamma)>0$ for any $\gamma>0$ and $\lim_{\gamma\to 1}\gamma'(\gamma)=1$.

Further, by our assumption on the error probability and on the delay  (see Definition~\ref{def:delaiscode}), we have for any message~$m$ 
 $$\pr_m(\cE^c_m\cap\{\tau-\nu\leq d_B-1\})\to 1 \quad (B\to \infty),$$ where $\cE^c_m$ denotes the successful decoding event. This implies that for any message~$m$
 $$\pr_m(0\leq \tau-\nu\leq d_B-1\})\to 1\quad (B\to \infty),$$ since the error probability is bounded away from zero whenever $\tau<\nu$.
 
 It then follows that 
\begin{align}\label{eq:tcond}\pr_m(\{0\leq \tau-\nu\leq d_B-1\}\cap \cI_{\gamma'})=1-o(1)\quad (B\to \infty).\end{align}
Hence, since $\nu$ is uniformly distributed over $\{1,2,\ldots,A+n-1\}$, for $B$ large enough we have
$$\pr_m(\{0\leq \tau-t\leq d_B-1\}\cap \cI_{\gamma'}|\nu=t)>0$$
for at least $(1-o(1))A$ values of $t\in \{1,2,\ldots, A\}$.
Now, conditioned on $\{\nu=t\}$, if event $$\{0\leq \tau-t\leq d_B-1\}\cap \cI_{\gamma'}$$ happens (\ie, with non-zero probability), then necessarily the period $\{t, t+1,\ldots, t+d_B-1\}$ contains at least $\gamma' n_B^*$ sampling times---here we use the fact that $\cS$ is non-randomized.

It then follows that 
\begin{align}\label{sam}
|\cS|\geq \left\lfloor \frac{(1-o(1)) A }{d_B}\right\rfloor  \gamma'\cdot n_B^*\,.
\end{align}

Now if 
\begin{align}\label{fineq}\rho d_B\leq n_B^*(1-\varepsilon)
\end{align}
for some arbitrary fixed $0<\varepsilon<1$, then  
\begin{align}
\left\lfloor \frac{(1-o(1)) A }{d_B}\right\rfloor  \gamma' n_B^*\geq \frac{ (1-o(1))\gamma'}{1-\varepsilon} \rho A(1-o(1))\label{sami}
\end{align}
as $B\to \infty$.

Hence, by taking $\gamma'$ and hence $\gamma$ close enough to $1$ and by taking $B$ large enough $$ (1-o(1)) \gamma'/(1-\varepsilon)>1.$$ Therefore,  if \eqref{fineq} holds, from \eqref{sam} and \eqref{sami} we get 
\begin{align}\label{bs}
|\cS|\geq \rho(1+\varepsilon')A
\end{align} for $B$ large enough and some $\varepsilon'>0$ such that $\varepsilon'\to 0$ as $\varepsilon \to 0$. Inequality \eqref{bs} implies that the sampling constraint is violated,  as we now show. 

Fix an arbitrary $0<\varepsilon''<1$.
For an arbitrary integer $1\leq k\leq A+n-1$ and any message~$m$
\begin{align}\label{eq3}
\pr_m(S_\tau &\geq  \rho \tau(1+\varepsilon''))\nonumber\\
&\geq \pr_m(S_\tau \geq \rho \tau(1+\varepsilon'')|\tau\geq k)\pr_m(\tau\geq k)\nonumber\\
&\geq \pr_m(S_k\geq \rho (A+n-1)(1+\varepsilon'')|\tau\geq k)\pr_m(\tau\geq k)\nonumber\\
&= \pr_m(S_k\geq (1+\varepsilon'')\rho A(1+o(1)))\pr_m(\tau\geq k)
\end{align} 
where for the second inequality we used the fact that the sampling times $S_1,S_2,\ldots$ are non-decreasing and the fact that $\tau\leq A+n-1$.  We now show that both terms $$\pr_m(S_k\geq (1+\varepsilon'')\rho A(1+o(1)))$$ and $$\pr_m(\tau\geq k)$$ are bounded away from zero in the limit $B\to \infty$, for an appropriate choice of $k$. This, by \eqref{eq3}, implies that 
$$\liminf_{B\to \infty}\pr_m(S_\tau \geq  \rho \tau(1+\varepsilon''))>0,$$
\ie, that sampling frequency $\rho$ is not achievable whenever \eqref{fineq} holds.  In other words, to achieve a sampling frequency $\rho$ it is necessary that delay and codeword length satisfy
$$d_B\geq \frac{n_B^*}{\rho}(1-o(1))\,.$$

Let
\begin{align}\label{eq8}
k=(1+2\varepsilon'')\rho A\,.
\end{align}
 Since $S_k\geq k$, 
\begin{align*}
S_k& 
\geq (1+2\varepsilon'')\rho A\,,
\end{align*}
and so by choosing $\varepsilon''>0$ small enough we get
\begin{align}\label{eq1}
 \pr(S_k\geq (1+\varepsilon'')\rho A(1+o(1)))=1
 \end{align}
 for $B$ large enough.
 
Since $\cC_B$ achieves (maximum) error probability $\leq \varepsilon_B$ we have for any message $m$
\begin{align}\label{eq5}
\varepsilon_B&\geq \pr_m(\E)\nonumber\\
&\geq \pr_m(\E|\tau<k,\nu \geq k)\pr_m(\tau<k,\nu\geq k)\nonumber\\
&\geq\frac{1}{2} \pr_m(\tau<k |\nu \geq k)\pr(\nu\geq k)\nonumber\\
&= \frac{1}{2}\pr_\star(\tau<k)\pr(\nu\geq k)\,.
\end{align}
For the third inequality in \eqref{eq5} note that event $\{\tau<k,\nu \geq k\}$ means that the decoder declares a message before the actual message even starts being sent. In this case, the error probability is at least $1/2$, since a message set always contains at least two messages (see Section~\ref{moper}). For the last equality in \eqref{eq5}, note that event $\{\tau\geq k\}$ depends only on $Y_1^k$, which are i.i.d.~$\sim Q_\star$ when conditioned on $\{\nu>k\}$---$\pr_\star$ denotes the output distribution under pure noise, \ie, when $Y_1^{A+n-1}$ is an i.i.d. $Q_\star$ random sequence.  Repeating this last change of measure argument we get
\begin{align}
\pr_m(\tau\geq k)&\geq \pr_m(\tau\geq k|\nu> k)\pr(\nu> k)\nonumber\\
&= \pr_\star(\tau\geq k)\pr(\nu> k)\nonumber\\
&\geq (1-2\varepsilon_B/\pr(\nu\geq k))\pr(\nu> k)\nonumber\\
&= (1-o(1))\pr(\nu> k)\nonumber\\
&= (1-\rho(1+2\varepsilon''))(1-o(1))\quad B\to \infty\,.\label{eq9}
\end{align}
The second inequality follows from \eqref{eq5}. For the second and third equality in \eqref{eq9} we use the fact that $\nu$ is uniformly distributed over $\{1,2,\ldots,A\}$, hence by \eqref{eq8} $$\pr(\nu> k)=(A-k)/A=1-(1-\rho(1+2\varepsilon''))>0\,.$$

Since $\rho(1+\varepsilon'')>0$, we have $\liminf_{B\to \infty}\pr_m(\tau\geq k)>0$ by \eqref{eq9}, yielding the desired claim.

Finally, to see that randomized sampling strategies cannot
achieve a better sampling frequency,
note that a randomized sampling strategy can be viewed as a probability distribution over deterministic sampling strategies. Therefore, because the
previous analysis holds for any deterministic sampling strategy,
it must also hold
for randomized sampling strategies rules.

\noindent{\it{Case $\star\notin \cX$:}}
Pick some arbitrary $\varepsilon>0$ and consider a code $\cC_B$ with length $n_B$ codewords that achieves rate per unit cost $\cC(\beta,\rho)-\varepsilon_B$, error probability $\leq \varepsilon_B$, delay $d_B=d(\cC_B,\varepsilon_B)$, and sampling frequency $\rho({\cal{C}}_B,\varepsilon_B)\leq \rho +\varepsilon_B$ for some  $\varepsilon_B\overset{B\to \infty}{\longrightarrow}0$. As in the previous case, without loss of optimality the sampling strategy $\cS$ is supposed to be non-randomized.

Because $\star\notin \cX$, we have $k(x)>0$ for any $x\in \cX$ and therefore to achieve the full-sampling asynchronous capacity per unit cost it is necessary that the codeword length remains essentially the same as under full sampling. More specifically, we must have
\begin{align}{n}_B\leq n_B'(1+\eta(\varepsilon))\quad B\to \infty\label{bbn}
\end{align}
for some $\eta(\varepsilon)\to 0$ as $\varepsilon\to0$, where
$n_B'$ denotes the number of sampled codeword positions---recall that codeword positions are the positions from time $\sigma$ up to time $\sigma+n_B-1$. 
Note that this is in contrast with the case $\star\in \cX$, where the codeword transmission 
duration can be expanded by transmitting $\star$  at no cost.

Proceeding as for the case $\star \in \cX$, 
we have
\begin{align}\label{eqhw}
\pr_m(\{0\leq \tau-t\leq d_B-1\}\cap \cA_{\gamma(\varepsilon)})=1-o(1)
\end{align}

$$\pr_m(\{0\leq \tau-t\leq d_B-1\}\cap \cA_{\gamma(\varepsilon)}|\nu=t)>0$$
for any $t\in \cB$ where  ${\cal{B}}$ is a certain subset of $\{1,2,\ldots, A\}$ with $|{\cal{B}}|=(1-o(1))A$. This means that for any $t\in {\cal{B}}$ the decoder samples a ``block'' $b\subseteq \cS$ of cardinality at least $\gamma n_B$ over the period $[t,t+1,\ldots,t+d_B-1]$. 
Moreover, if we denote by $i(b)$ and $f(b)$ the time position within $\{1,2,\ldots,A+n-1\}$ of the first and the last element of $b$, respectively, then for each block we have $f(b)-i(b)\leq n_B$.

Because of the sampling constraint, there are at most 
$$N=\frac{\rho A (1+o(1))}{\gamma n_B}$$
distinct blocks of size $\gamma n_B$. This implies that $d_B$ should satisfy
$$d_B\geq (\gamma n_B/\rho + \gamma n_B)(1-o(1)),$$
as we now show. Intuitively, the reason the delay must satisfy this
bound is that because the codewords must now be blocks of symbols,
the receiver might as well sample in blocks of $n_B$ symbols. Then,
the sampling constraint means that, on average, the gap between sampled
blocks grows like $n_B/\rho$. However, if the message arrives at a time $\nu$
close to the beginning of a block, then in addition to waiting until the next block,
the message must wait for most of the current block before being transmitted---close to capacity we cannot afford to miss a portion of the codeword other than negligible.
Therefore, the delay must grow as $n_B/\rho+n_B$. We formalize this reasoning
below.

Suppose, by way of contradiction, that 
\begin{align}\label{eqb}d_B\leq \gamma n_B(1+1/\rho )(1-\varepsilon)\end{align}
for some $\varepsilon>0$, and assume for the moment that each block $b_{(t)}$ is composed of  $\gamma n_B(1+o(1))$ elements and that there are least $N(1-o(1))$ distinct blocks.

Define the ``occupation'' slot of a block as $\gamma n_B$ plus the time interval until the next block. The average occupation slot per block is thus
$$\frac{A}{N}=\frac{\gamma n_B}{\rho}(1-o(1))\,.$$
Hence, for any $\varepsilon'>0$ there is a set of at least $\eta N$ occupation slots each of size at most
$$\frac{\gamma n_B}{\rho}(1+\varepsilon')$$
where $\eta=\eta(\varepsilon')>0$ for any $\varepsilon'>0$. Consider such a set of occupation slots for some $\varepsilon'>0$ which is specified later, let $b$ be a block belonging to one such slots, and let $b'$ denote the block coming after $b$.\footnote{For reasons that will soon be obvious, $b$ should not be the right most block within the set.} 
Denote by $i(b)$ and $f(b)$ the time position within $\{1,2,\ldots,A+n-1\}$ of the first and the last element of $b$, respectively.
Then for $B$ large enough $$f(b')-i(b)\geq \gamma n_B(1+\frac{1 }{\rho}(1+\varepsilon'))$$
and therefore by taking $\varepsilon'>0$ small enough we get $$f(b')-i(b)>\eta' n_B$$
 by \eqref{eqb}, where $\eta'=\eta'(\varepsilon,\varepsilon')>0$ for any $\varepsilon>0$ and  $\varepsilon'>0$.
It then follows that for any $t\in (i(b),i(b)+\eta'n_B]$, the interval $[t,t+1,\ldots,t+d_B-1]$ contains neither blocks $b$ and $b'$ completely. Therefore, conditioned on $\nu\in (i(b),i(b)+\eta'n_B]$, event
$$\{\tau-\nu\leq d_B-1\}\cap \cI_\gamma$$ does not happen. It then follows that if \eqref{eqb} holds for some $\varepsilon>0$, then 
$$\limsup_{B\to \infty}\pr_m (\{\tau-\nu\leq d_B-1\}\cap \cI_\gamma)\leq 1-\eta \eta'<1$$
which contradicts \eqref{eqhw}.

The above argument assumes that there are $N$ disjoint blocks of size $\gamma n_B$. If there are fewer and possibly larger blocks, the arguments easily extend by defining the blocks $b$ as any subset of $\cS$ such that $f(b)-i(b)\leq n_B$.
 \end{IEEEproof}

\begin{IEEEproof}[Proof of Theorem~\ref{thm:adaptive}]
We show that 
$\bm{C}(\beta,\rho)=\bm{C}(\beta,1)$ for any $\beta>0$ and $0<\rho\leq 1$ and that $\bm{C}(\beta,1)$ can be achieved with codes $\{\cC_B\}$ with delay $d(\cC_B,\varepsilon_B)=n^*_B(1+o(1))$ as $B\to \infty$.
 
Let $P$ be the distribution achieving $\bm{C}(\beta,1)$ (see
Theorem~\ref{fullsampling}). We generate $2^B$ codewords of
length $$n-\log(n)$$ as in
the proof of Theorem~\ref{thm:nonadaptive} according to
distribution $P$. 
Each codeword starts with  a common preamble that consists
of $\log(n)$ repetitions of a symbol $x$ such that
$Q(\cdot|x)\ne Q(\cdot|\star)$.

For the proposed  asymptotically optimal sampling/decoding strategy, it is convenient to
 introduce the following notation. Let $\tilde{Y}_a^b$ denote the  
random vector obtained by extracting the components of the output process $Y_t$ at $t \in [a,b]$
of the form $t = \lceil j/\f \rceil$ for non-negative integer $j$. 
Notice that, for any $t \geq \ell$ and $\ell \gg 1$,  $\tilde{Y}_{t - \ell + 1}^t$ contains $\approx \rho \ell$ samples.
 
The strategy starts in the 
sparse mode, taking samples at times $S_j = \lceil j/\f\rceil$, $j=1,2,\ldots$. 
At each $j$, the receiver computes the empirical distribution (or type)
$$\hat{P}_j=\hat{P}_{\tilde{Y}^{S_j}_{S_j- \log (n)+1}}$$
of the sampled output in the most recent window of length $\log(n)$. 

If the probability of this type under pure noise is large enough, \ie, if 
$$\pr_\star(\cT_{\hat{P}_j})>\frac{1}{n^2}\,,$$
the mode is kept unchanged and we repeat this test at the next round $j+1$.

Instead, if  $$\pr_\star(\cT_{\hat{P_j}})\leq \frac{1}{n^2}\,,$$
then we switch to the dense sampling mode, taking samples 
continuously for at most $n$ time steps.  
At each of these steps the receiver applies the same sequential typicality
decoder as in the proof of Theorem~\ref{thm:nonadaptive},
based on the past $n-\log n$ output samples. If no codeword is typical with the channel outputs 
after these $n$ times steps, sampling is switched back to the sparse mode.

We  compute the error
probability of the above scheme, its relative number of samples, and its delay. 

For the error probability, a similar analysis as for the
non-adaptive case in the proof of 
Theorem \ref{thm:nonadaptive} still applies, with $\f n$ being
replaced by $n-\log n$. In particular, after fixing the
ratio $B/n$ and thereby imposing a delay linear in $B$, equation
\eqref{eqmaster} holds with $\f=1$. 

For the relative number of samples, we now show that
\begin{align}
\pr_{m}(\tau/S_\tau\geq
\f+\varepsilon_B)\overset{n\to \infty}{\longrightarrow} 0
\end{align}
with $\varepsilon_B=1/\poly(B) $ from which we then conclude that $\cC(\beta,\rho)\geq
\cC(\beta,1)$. To do this, it is convenient to
introduce $Z_i$, $1\leq i\leq A+n-1$, which is equal to one if at
time $i$ the receiver switches to the dense mode and samples the next $n$ channel
outputs and equal to zero otherwise. Then it follows that
\begin{align}\label{nrsam}
\tau \leq \f S_\tau + n \sum_{i=1}^{S_\tau} Z_i\,.
\end{align}
To see this, note that the number of samples involved in the
sparse mode is at most $\f S_\tau$ and that the number of
samples involved in the dense mode is at most $n\sum_{i=1}^{S_\tau} Z_i$ (it is
actually equal to $n \sum_{i=1}^{S_\tau} Z_i$ if we ignore
the boundary discrepancies that we cannot sample beyond time $A+n-1$).

From \eqref{nrsam}
\begin{align}\label{sramdet}
\pr(\tau/S_\tau\geq \f&+\varepsilon)\leq \pr(n\sum_{i=1}^{S_\tau}Z_i\geq
S_\tau\varepsilon)\notag\\
&\leq \pr(n\sum_{i=1}^{S_\tau}Z_i\geq
S_\tau\varepsilon,\nu\leq S_\tau\leq
\nu+2n-2)\notag\\
&+\pr(S_\tau<\nu \:\text{or}\:S_\tau> \nu+2n-2)\,.
\end{align}
We now show that the right-hand side of the second inequality in  \eqref{sramdet} vanishes as $B\to \infty$.

For the first term on the right-hand side of the second inequality in \eqref{sramdet},
since the $Z_i$'s are nonnegative
\begin{align}\label{dex}
\pr(n\sum_{i=1}^{S_\tau}Z_i\geq
S_\tau&\varepsilon;\nu\leq S_\tau\leq
\nu+2n-2)\nonumber \\
&\leq \pr(n\sum_{i=1}^{\nu+2n-2}Z_i\geq
\nu\varepsilon)\,. 
\end{align}
Now, conditioned on $\nu=t$, the $Z_i$'s, $1\leq i\leq t-1$, are binary random variables distributed according to pure noise. Hence,
\begin{align}\label{primamon}
\pr\Big(n\sum_{i=1}^{t+2n-2}&Z_i\geq
t\varepsilon|\nu=t\Big)\nonumber\\
&\leq \pr_\star\Big(n\sum_{i=1}^{t-1}Z_i\geq t\varepsilon -(2n-1)\Big)\nonumber\\
&\leq \frac{t-1}{(t\varepsilon-(2n-1)-(t-1)/n^2)^2}\nonumber \\
&=o(1) \quad (t\to \infty)
\end{align}
where the second inequality follows from Chebyshev's inequality
and by noting that for $1\leq i\leq t-1$ we have $$\text{Var}(Z_i)\leq \ex Z_i\leq 1/n^2$$ since the variance of a
Bernoulli random variable is at most its mean which, in turn, is at most $1/n^2$.

Therefore,
\begin{align}\label{mim2}
\pr\Big(n&\sum_{i=1}^{\nu+2n-1}Z_i\geq
\nu\varepsilon)\nonumber \\
&\leq \pr(\nu\leq \sqrt{A}\Big)\nonumber \\
&+\frac{1}{A}\sum_{t=\sqrt{A}+1}^{A}\pr(n\sum_{i=1}^{\nu+2n-1}Z_i\geq
\nu\varepsilon|\nu=t)\nonumber \\
&=o(1)\quad (B\to\infty)
\end{align}
where the last equality follows from \eqref{primamon} and the fact that $\nu$ is uniformly distributed over $\{1,2,\ldots, A=e^{\beta B}\}$. From \eqref{dex} and \eqref{mim2} we get
$$\pr(n\sum_{i=1}^{S_\tau}Z_i\geq
S_\tau\varepsilon;\nu\leq S_\tau\leq
\nu+2n-2)=o(1) $$
as $B\to \infty$.

We now show that 
\begin{align}\label{igw}
\pr(S_\tau<\nu \:\text{or}\:S_\tau\geq  \nu+2n-1)\to 0\quad (B\to \infty)\,.
\end{align}
That $\pr(S_\tau<\nu )\to 0$ follows from the fact that $\pr_m(\E_2)\to 0$ where
$\E_2$ is defined in the proof of
Theorem~\ref{thm:nonadaptive}. That $\pr(S_\tau\geq \nu+2n-1)\to
0$ follows from the fact that with probability tending to one
the sampling strategy will changes mode over the transmitted
codeword and that the typicality decoder will make a decision up to time
$\nu+n-1$ with probability tending to one. This last argument can also be used for the delay to show that $d(\cC_B,\varepsilon_B)=n(1+o(1))$ for some $\varepsilon_B\to 0$. 

Finally, by optimizing the input distribution to minimize delay (see paragraph after Theorem~\ref{fullsampling}) we deduce that $\cC(\beta,1)=\cC(\beta,\rho)$ and that the capacity per unit cost is achievable with delay $n^*_B(1+o(1))$.
\end{IEEEproof}

 We end this section with a few words concerning the Remark~\ref{rem1} at the end of Section~\ref{revres}. To prove the claim it suffices to slightly modify the achievability schemes yielding Theorems~\ref{thm:nonadaptive} and \ref{thm:adaptive} to make them universal at the decoder. 
  
 The first modification is needed to estimate the pure noise distribution $Q_\star$ with a negligible fraction of channel outputs. An estimate of this distribution is obtained by sampling the first $\sqrt{A}$ output symbols. At the end of this estimation phase, the receiver declares the pure noise distribution as being equal to $\hat{P}_{Y_1^{\sqrt{A}}}$. Note that since $\nu$ is uniformly distributed over $\{1,2,\ldots,A\}$ we have $$\pr(||\hat{P}_{Y_1^{\sqrt{A}}}-Q_\star||_1\geq \varepsilon_B)\to 1$$
as $B\to \infty$, for some $\varepsilon_B\to 0$. Note also that this estimation phase requires a negligible amount of sampling, {\it{i.e.}}, sublinear in $A$.

The second modification concerns the typicality decoder which is replaced by an MMI (Maximum Mutual Information) decoder (see \cite[Chapter 2]{CK}).

It is straightforward to verify that the modified schemes indeed achieve the asynchronous capacity per unit cost. The formal arguments are similar to those used in \cite[Proof of Theorem 2]{tchamkerten2009communication} (see also \cite[Theorem 3]{6365818} which proves the claim under full sampling and unit input cost) and are thus omitted.

\bibliographystyle{IEEEtran}
\bibliography{../../common_files/bibiog}

\begin{thebibliography}{10}
\providecommand{\url}[1]{#1}
\csname url@samestyle\endcsname
\providecommand{\newblock}{\relax}
\providecommand{\bibinfo}[2]{#2}
\providecommand{\BIBentrySTDinterwordspacing}{\spaceskip=0pt\relax}
\providecommand{\BIBentryALTinterwordstretchfactor}{4}
\providecommand{\BIBentryALTinterwordspacing}{\spaceskip=\fontdimen2\font plus
\BIBentryALTinterwordstretchfactor\fontdimen3\font minus
  \fontdimen4\font\relax}
\providecommand{\BIBforeignlanguage}[2]{{%
\expandafter\ifx\csname l@#1\endcsname\relax
\typeout{** WARNING: IEEEtran.bst: No hyphenation pattern has been}%
\typeout{** loaded for the language `#1'. Using the pattern for}%
\typeout{** the default language instead.}%
\else
\language=\csname l@#1\endcsname
\fi
#2}}
\providecommand{\BIBdecl}{\relax}
\BIBdecl

\bibitem{6397617}
V.~Chandar, A.~Tchamkerten, and D.~Tse, ``Asynchronous capacity per unit
  cost,'' \emph{Information Theory, IEEE Transactions on}, vol.~59, no.~3, pp.
  1213 --1226, march 2013.

\bibitem{chandar2008optimal}
V.~Chandar, A.~Tchamkerten, and G.~Wornell, ``Optimal sequential frame
  synchronization,'' \emph{Information Theory, IEEE Transactions on}, vol.~54,
  no.~8, pp. 3725--3728, 2008.

\bibitem{tchamkerten2009communication}
A.~Tchamkerten, V.~Chandar, and G.~Wornell, ``Communication under strong
  asynchronism,'' \emph{Information Theory, IEEE Transactions on}, vol.~55,
  no.~10, pp. 4508--4528, 2009.

\bibitem{6352910}
A.~Tchamkerten, V.~Chandar, and G.~W. Wornell, ``Asynchronous communication:
  Capacity bounds and suboptimality of training,'' \emph{Information Theory,
  IEEE Transactions on}, vol.~59, no.~3, pp. 1227 --1255, march 2013.

\bibitem{6365818}
Y.~Polyanskiy, ``Asynchronous communication: Exact synchronization,
  universality, and dispersion,'' \emph{Information Theory, IEEE Transactions
  on}, vol.~59, no.~3, pp. 1256 --1270, march 2013.

\bibitem{wang2011error}
D.~Wang, V.~Chandar, S.~Chung, and G.~Wornell, ``Error exponents in
  asynchronous communication,'' in \emph{Information Theory Proceedings (ISIT),
  2011 IEEE International Symposium on}.\hskip 1em plus 0.5em minus 0.4em\relax
  IEEE, 2011, pp. 1071--1075.

\bibitem{shomorony2012bounds}
I.~Shomorony, R.~Etkin, F.~Parvaresh, and A.~Avestimehr, ``Bounds on the
  minimum energy-per-bit for bursty traffic in diamond networks,'' in
  \emph{Information Theory Proceedings (ISIT), 2012 IEEE International
  Symposium on}.\hskip 1em plus 0.5em minus 0.4em\relax IEEE, 2012, pp.
  801--805.

\bibitem{6283479}
M.~Khoshnevisan and J.~Laneman, ``Achievable rates for intermittent
  communication,'' in \emph{Information Theory Proceedings (ISIT), 2012 IEEE
  International Symposium on}, july 2012, pp. 1346 --1350.

\bibitem{D}
R.~L. Dobrushin, ``Asymptotic bounds on the probability of error for the
  transmission of messages over a memoryless channel using feedback,''
  \emph{Probl. Kibern.}, vol.~8, pp. 161--168, 1963.

\bibitem{CK}
I.~Csisz\`ar and J.~K\"orner, \emph{Information Theory: Coding Theorems for
  Discrete Memoryless Channels}.\hskip 1em plus 0.5em minus 0.4em\relax New
  York: Cambridge University Press, 2011.

\bibitem{CLRS}
T.~H. Cormen, C.~E. Leiserson, R.~L. Rivest, and C.~Stein, \emph{Introduction
  to Algorithms, 2nd edition}.\hskip 1em plus 0.5em minus 0.4em\relax {MIT}
  Press, McGraw-Hill Book Company, 2000.

\bibitem{Gr}
G.~Grimmett and D.~Stirzaker, \emph{One Thousand Exercises in
  Probability}.\hskip 1em plus 0.5em minus 0.4em\relax New York: Oxford
  University Press, 2001.

\end{thebibliography}

\end{document}